\documentclass[superscriptaddress]{revtex4}
\usepackage{epsfig}
\def\ii{\'{\i}}
\def\beq{\begin{equation}}
\def\eeq{\end{equation}}
\def\beqa{\begin{eqnarray}}
\def\eeqa{\end{eqnarray}}
\def\ban{\begin{eqnarray*}}
\def\ean{\end{eqnarray*}}
\def\bi{\begin{itemize}}
\def\ei{\end{itemize}}

\def\d{\mbox{d}}

\begin{document}

\title{Relativistic Mean Field Approximation in a Density Dependent
Parametrization Model at Finite Temperature}

\author{S.S. Avancini}
\affiliation{Depto de F\'{\i}sica - CFM - Universidade Federal de Santa
Catarina  Florian\'opolis - SC - CP. 476 - CEP 88.040 - 900 - Brazil}
\author{M.E. Bracco}
\affiliation{Instituto de F\ii sica - Universidade do Estado do Rio de Janeiro - Rua S\~ao Francisco Xavier 524 - Maracan\~a - CEP 20559-900 - Rio de Janeiro - RJ - Brazil}
\author{M. Chiapparini}
\affiliation{Instituto de F\ii sica - Universidade do Estado do Rio de Janeiro - Rua S\~ao Francisco Xavier 524 - Maracan\~a - CEP 20559-900 - Rio de Janeiro - RJ - Brazil}
\author{D.P.Menezes}
\affiliation{Depto de F\'{\i}sica - CFM - Universidade Federal de Santa
Catarina  Florian\'opolis - SC - CP. 476 - CEP 88.040 - 900 - Brazil}

\begin{abstract}
In this work we calculate the equation of state of nuclear matter
for different proton fractions at zero and finite temperature
within the Thomas Fermi approach considering
three different parameter sets: the well-known NL3 and TM1 and a
density dependent parametrization proposed by Typel and Wolter.
The main differences are
outlined and the consequences of imposing beta-stability in these
models are discussed.
\end{abstract}

\maketitle

\vspace{0.50cm}
PACS number(s): {\bf 21.65.+f, 21.30.-x, 25.70.-z}
\vspace{0.50cm}

\section{Introduction and formalism}

Understanding the properties of nuclear matter at both, normal and high
densities, is of
crucial importance in explaining the appearance of neutron stars after the
supernova explosion and the formation of transiron elements in nuclear
reactions.

One of the most popular relativistic models is the non-linear
Walecka model \cite{bb,ring}, which can be used in order to obtain
different equations of state (EOS) as far as different parameter
sets are employed. In this work we investigate the consequences in
the EOS when the parametrizations of the well known NL3 model
\cite{nl3}, which is a good parametrization in describing finite
nuclei properties, are changed to the density dependent one
proposed in \cite{tw}. The new parametrization is determined by fitting
several nuclear matter bulk properties and also some finite nuclei.
Both models were investigated considering two types of
proton fractions: fixed ones and those arising when
$\beta$-equilibrium is incorporated. Some considerations are also
done in relation to the TM1 parameter set \cite{tm1}. The
extension of the density dependent parametrization to finite
temperature EOS is also investigated. This extension has been
partially studied in \cite{ditoro} for symmetric nuclear matter
only. In nuclear collisions involving stable or radioactive
neutron rich nuclei, in experiments yielding nuclear
multifragmentation, in protoneutron stars, among inumerous
examples, the resulting matter is known to carry a reasonable
amount of isospin asymmetry. Hence, in the present work, a more
complete and detailed study is performed in order to account for
asymmetric matter as well.

We start from the lagrangian density of the relativistic
non-linear model, adapted in order to accommodate the NL3, TM1
forces and the density dependent meson-nucleon coupling constants
\cite{tw}: \beqa {\cal L}&=&\bar \psi\left[
\gamma_\mu\left(i\partial^{\mu}-\Gamma_\omega \omega^{\mu}-
\frac{\Gamma_{\rho}}{2} \vec{\tau} \cdot \vec{\rho}^\mu \right)
-(M-\Gamma_\sigma \sigma)\right]\psi \nonumber \\
&&+\frac{1}{2}(\partial_{\mu}\sigma\partial^{\mu}\sigma
-m_\sigma^2 \sigma^2)
- \frac{1}{3!}\kappa \sigma^3 -\frac{1}{4!}\lambda\sigma^4
\nonumber \\
&& -\frac{1}{4}\omega_{\mu\nu}\omega^{\mu\nu}+\frac{1}{2}
m_\omega^2 \omega_{\mu}\omega^{\mu} + \frac{1}{4!} \xi \Gamma_{\omega}^4
(\omega_{\mu}\omega^{\mu})^2 \nonumber \\
&&-\frac{1}{4}\vec\rho_{\mu\nu}\cdot\vec \rho^{\mu\nu}+\frac{1}{2}
m_\rho^2 \vec\rho_{\mu}\cdot \vec \rho^{\mu},
\label{lag1}
\eeqa
where $\sigma$, $\omega^\mu$ and $\vec{\rho^\mu}$ are the
scalar-isoscalar, vector-isoscalar and vector-isovector meson fields
respectively,
$\omega_{\mu\nu}=\partial_{\mu}\omega_{\nu}-\partial_{\nu}\omega_{\mu}$,
and
$\vec \rho_{\mu\nu}=\partial_{\mu}\vec \rho_{\nu}-\partial_{\nu} \vec \rho_{\mu}
- \Gamma_\rho (\vec \rho_\mu \times \vec \rho_\nu)$. Besides this, $M$ is the nucleon mass,
$m_\sigma$, $m_\omega$, $m_\rho$ are the masses of the mesons and
 $\Gamma_\sigma$, $\Gamma_\omega$, $\Gamma_\rho$ are the nucleon-meson coupling constants.
 $\kappa$, $\lambda$ and $\xi$ are the self-interacting scalar
and vector coupling constants. In this work we investigate the
differences arising from three parameter sets, namely NL3
\cite{nl3}, TM1 \cite{tm1} and TW \cite{tw}. In the first two
cases, $\Gamma_{\sigma}$, $\Gamma_{\omega}$ and $\Gamma_{\rho}$
are the usual $g_{\sigma}$, $g_{\omega}$ and $g_{\rho}$. In the second case,
the density dependent coupling constants are adjusted in order to reproduce
some of the nuclear matter bulk properties, using the parametrization
given in \cite{tw}. Other possibilities
for these parameters are also found in the literature \cite{ditoro}.
In the TW parametrization the meson self-coupling constants $\kappa$,
$\lambda$ and $\xi$ are zero. The nuclear matter bulk properties described
by these three parameter sets are displayed in table 1.

\indent From the Euler-Lagrange equations we obtain the field equations of motion
in the mean field appro\-xi\-ma\-tion for infinite matter, where the meson fields are
replaced by their expectation values. In this approximation, the expectation
value of the $\sigma$, $\omega$ and $\rho$ meson fields are called $\phi_0$,
$V_0$ and $b_0$ respectively. The coupled equations read
\begin{eqnarray}
m_{\sigma}^2\phi_0 +\frac{1}{2}\kappa \phi^2_0 +\frac{1}{3!}
\lambda\phi^3_0
- \Gamma_{\sigma} \rho_s &=&0, \label{phi} \\
m_{\omega}^2 V_0 + \frac{1}{3!} \xi \Gamma_{\omega}^4 V_0^3
- \Gamma_{\omega} \rho&=&0, \label{V0}\\
m_\rho^2 b_0 -\frac{\Gamma_\rho}{2} \rho_3&=&0, \label{b0}\\
\left[ i\gamma^{\mu}\partial_\mu -\gamma_0\left( \Gamma_{\omega}
V_0+\frac{\Gamma_{\rho}}{2} \tau_3 b_0 +{\Sigma}^{R}_0 \right)
-M^{\ast} \right] \psi&=&0, \label{dirac}
\end{eqnarray}
where the {\it rearrangement} term $\Sigma^{R}_0$ is given by
\begin{equation}
\Sigma^{R}_0=\frac{\partial \Gamma_{\omega}}{\partial \rho} \rho V_0 +
\frac{\partial \Gamma_\rho}{\partial \rho} \rho_3 \frac{b_0}{2} -
\frac{\partial \Gamma_{\sigma}}{\partial \rho} \rho_s \phi_0,  \label{rearr}
\end{equation}
and the scalar and baryonic densities are defined as
\begin{eqnarray}
\rho_s&=&\langle\bar \psi \psi\rangle, \\
\rho&=&\langle\bar \psi \gamma^0 \psi\rangle, \\
\rho_3&=&\langle\bar \psi \gamma^0 \tau_3\psi\rangle .
\end{eqnarray}

In the following discussion we consider nuclear matter in the the mean-field
approximation only for the TW parameter set. Due to translational and
rotational invariance the lagrangian density reduces to
\begin{eqnarray}
{\cal L}_{MFT}&=&\bar \psi\left[ i\gamma_\mu
\partial^{\mu}-\gamma_0\Gamma_{\omega} V_0-
\gamma_0\frac{\Gamma_{\rho}}{2} \tau_3 b_0 -(M-\Gamma_{\sigma}
\phi_0)\right]\psi \nonumber \\
&&
-\frac{1}{2} m_{\sigma}^2 \phi_0^2 + \frac{1}{2} m_{\omega}^2 V_{0}^2
+\frac{1}{2} m_\rho^2 b_{0}^{2}. \label{lag2}
\end{eqnarray}

The conserved energy-momentum tensor can be derived in the usual
fashion \cite{sw}:
\begin{equation}
{\cal T}_{MFT}^{\mu\nu}=\bar \psi i\gamma^\mu \partial^{\nu}\psi
-g^{\mu\nu}\left[ \frac{1}{2} m_{\sigma}^2 \phi_0^2 - \frac{1}{2} m_{\omega}^2
V_{0}^2 -\frac{1}{2} m_\rho^2 b_{0}^{2}-\bar\psi \gamma_0
\Sigma^{R}_0 \psi \right]  \label{tem} .
\end{equation}
Note that the rearrangement term included above and defined in
eq.(\ref{rearr}) assures the
energy-momentum conservation, i.e., $\partial_\mu{\cal T}^{\mu\nu}=0$. From
the energy-momentum tensor one easily obtains the
hamiltonian operator:
\begin{eqnarray}
 {\cal H}_{MFT}=\int d^3 x ~{\cal T}_{MFT}^{00}&=&
 \int d^3 x ~ \psi^{\dagger}\left(-i\vec\alpha \cdot \nabla
 +\beta M^\ast + \Gamma_{\omega} V_0+
 \frac{\Gamma_{\rho}}{2} \tau_3 b_0\right) \psi  \nonumber \\
 &&+V\left(\frac{1}{2} m_{\sigma}^2 \phi_0^2
 - \frac{1}{2} m_{\omega}^2 V_{0}^2
-\frac{1}{2} m_\rho^2 b_{0}^{2}\right), \label{hamil}
\end{eqnarray}
where $M^{\ast}=M-\Gamma_{\sigma}\phi$ and $V$ is the volume of the
system. In the above equation the rearrangement term
cancels out. Notice that as a consequence, the energy density does not
carry the rearrangement term either and can be written in the semi-classical
Thomas-Fermi approximation as
\begin{eqnarray}
{\cal E}&=& 2 \sum_{i=p,n} \int \frac{\d^3p}{(2\pi)^3}
\sqrt{{\mathbf p}^2+{M^*}^2} \left(f_{i+}+f_{i-}\right)
+ \Gamma_{\omega} V_0 \rho +\frac{\Gamma_{\rho}}{2} b_0 \rho_3
\nonumber \\
&&+\frac{m_{\sigma}^2}{2} \phi_0^2-\frac{m_{\omega}^2}{2} V_0^2
-\frac{m_{\rho}^2}{2} b_0^2.\label{enerd}
\end{eqnarray}
Note that for the NL3 parametrization, the term $\kappa\phi_0^3/6 +\lambda\phi_0^4/24$
also appears in the energy density equation and for the TM1 these two terms
come together with $-\xi \Gamma_{\omega}^4 V_0^4/24$.
Following the notation in \cite{mp}, the thermodynamic potential
can be written as
\begin{equation}
\Omega= {\cal E} - T {\cal S} - \mu_p \rho_p - \mu_n \rho_n ,
\label{Omega1}
\end{equation}
where ${\cal S}$ is the entropy density of a classical Fermi gas, $T$ is the
temperature, $\mu_p$ ($\mu_n$) is the proton (neutron) chemical potential and
$\rho_p$ and $\rho_n$ are respectively the proton and neutron densities,
calculated in such a way that $\rho=\rho_p+\rho_n$. We have
\begin{equation}
\rho_i=2 \int\frac{\d^3p}{(2\pi)^3}(f_{i+}-f_{i-}), \quad i=p,n\; ,
\label{rhoi}
\end{equation}
where the distribution functions $f_{i+}$ and $f_{i-}$ for particles and
anti-particles have to be derived in order to make the thermodynamic
potential stationary for a system in equili\-bri\-um. After straightforward
substitutions, eq.(\ref{Omega1}) becomes
\begin{eqnarray}
\Omega&=&
2 \sum_{i=p,n} \int \frac{d^3p}{(2 \pi)^3} \sqrt{{\mathbf p}^2+{M^*}^2}
(f_{i+} + f_{i-})
+ \Gamma_{\omega} V_0 \rho +\frac{\Gamma_{\rho}}{2} b_0 \rho_3
+\frac{m_{\sigma}^2}{2} \phi_0^2-\frac{m_{\omega}^2}{2} V_0^2
-\frac{m_{\rho}^2}{2} b_0^2\nonumber \\
&&+2 T \sum_{i=p,n} \int \frac{d^3p}{(2 \pi)^3} \left(
f_{i+} \ln \left(\frac{f_{i+}}{1-f_{i+}}\right) + \ln ({1-f_{i+}}) +
f_{i-} \ln \left(\frac{f_{i-}}{1-f_{i-}}\right) + \ln ({1-f_{i-}}) \right)
\nonumber \\
&&-2 \sum_{i=p,n} \int \frac{d^3p}{(2 \pi)^3}\:\mu_i (f_{i+}-f_{i-}).
\label{Omega2}
\end{eqnarray}
For a complete demonstration of the above shown expressions
obtained in a
Thomas-Fermi approximation for the non-linear Walecka model, please refer to
\cite{mp}.
At this point, eq.(\ref{Omega2}) is minimized in terms of the distribution
functions for fixed meson fields, i.e.,
\begin{equation}
\left. \frac{\partial \Omega}{\partial f_{i+}}
\right |_ {{f_{i-},f_{j\pm}, \phi_0,V_0,b_0}} =0 \quad i \ne j.
\end{equation}
For the proton distribution function, the above calculation yields
\begin{equation}
E^{\ast}({\mathbf p}) + \Sigma^R_0 - \mu_p +
\Gamma_{\omega} V_0 + \frac{\Gamma_{\rho}}{2} b_0 = -T\:
\ln \left(\frac{f_{p+}}{1-f_{p+}}\right),
\end{equation}
where
$E^{\ast}({\mathbf p})=\sqrt{{\mathbf p}^2+{M^*}^2}$.
Similar equations, with some sign differences are obtained for the anti-proton,
neutron and anti-neutron distribution functions. The effective chemical
potentials are then defined as
\begin{eqnarray}
\mu^{\ast}_p&=&\mu_p-\Gamma_{\omega} V_0 -\frac{\Gamma_{\rho}}{2} b_0 -
\Sigma^{R}_0, \nonumber \\
\mu^{\ast}_n&=&\mu_n-\Gamma_{\omega} V_0 +\frac{\Gamma_{\rho}}{2} b_0 -
\Sigma^{R}_0 \label{efchem}
\end{eqnarray}
and the following equations for the distribution functions can be written:
\begin{equation}
f_{i\pm}=
\frac{1}{1+\exp[(E^{\ast}({\mathbf p}) \mp\mu^*_i)/T]}\;,
\quad i=p,n. \label{disfun}
\end{equation}
In the above calculation we have used
\[
\rho_s= 2 \sum_{i=p,n}
\int \frac{\d^3p}{(2\pi)^3}
\frac{M^*}{E^{\ast}({\mathbf p})}\left(f_{i+}+f_{i-}\right),
\]
and $\rho_3=\rho_p-\rho_n$.
The proton fraction is defined as $Y_p=\rho_p/\rho$.

Within the Thomas-Fermi approach the pressure becomes
\begin{eqnarray}
P&=&\frac{1}{3 \pi^2} \sum_{i=p,n}
\int \d p \frac{{\mathbf p}^4}{\sqrt{{\mathbf p}^2+{M^*}^2}} \left( f_{i+} + f_{i-}\right)
-\frac{m_{\sigma}^2}{2} \phi_0^2 \left( 1 + 2 \frac{\rho}{\Gamma_{\sigma}}
\frac{\partial
\Gamma_s}{\partial \rho} \right) \nonumber \\
&&+\frac{m_{\omega}^2}{2} V_0^2 \left( 1 + 2 \frac{\rho}{\Gamma_{\omega}}
\frac{\partial \Gamma_{\omega}}{\partial \rho} \right)
+\frac{m_{\rho}^2}{2} b_0^2 \left( 1 + 2 \frac{\rho}{\Gamma_{\rho}} \frac
{\partial \Gamma_{\rho}}{\partial \rho} \right).
\label{pressure}
\end{eqnarray}
In the NL3 model, the term $-\kappa\phi_0^3/6 -\lambda\phi_0^4/24$
is also present in (\ref{pressure}) and in the TM1 model these
terms are also accompanied by $\xi \Gamma_{\omega}^4 V_0^4/24$. It
is also important to stress that the thermodynamical consistency
which requires the equality of the pressure calculated from the
thermodynamical definition and from the energy-momentum tensor,
discussed in \cite{flw}, is also obeyed by the temperature
dependent TW model.

Another quantity of interest is the nuclear bulk symmetry energy discussed
in \cite{lkb}. It is usually defined as
\begin{equation}
{\cal E}_{sym} =\left. \frac{1}{2} \frac{\partial^2 {\cal E}}
{\partial \delta^2} \right|_{\delta=0},
\label{symen}
\end{equation}
with $\delta= \rho_3/\rho$ and which can be analytically rewritten
as
\begin{equation}
{\cal E}_{sym}=\left( \frac{k_F^2}{6 E^{\ast}(\mathbf p)}+ \frac{\Gamma_\rho^2}
{8 m_\rho^2} \right) \rho, \label{esym}
\end{equation}
where
\[
k_{Fp}=k_F(1+\delta)^{1/3},\qquad k_{Fn}=k_F(1-\delta)^{1/3},
\]
with $k_F=(1.5 \pi^2\rho)^{1/3}$.
The value and behavior of the symmetry energy at densities larger than nuclear
saturation density are still not well established. This quantity is important
in studies involving neutron stars and radioactive nuclei. In general,
relativistic and non-relativistic models give different predictions for the
symmetry energy. A comparison between the symmetry energies coming from the NL3
and the TW
models is also discussed in the present work.

\section{Considering $\beta$-stability}

At this point, we introduce the ideas of $\beta$ stability and
charge neutrality. In an ideal system of protons, neutrons,
electrons and muons in equilibrium, the particle levels are filled
in such a way that the $\beta$ decays are forbidden. In order to
study the conditions of $\beta$ equilibrium, one has to
incorporate leptonic degrees of freedom in the lagrangian density
of equation (\ref{lag1}), obtaining for the new lagrangian the
following expression:
\begin{equation}
{\cal L}_{lb}={\cal L} + {\cal L}_{leptons},
\label{lblag}
\end{equation}
where
\begin{equation}
{\cal L}_{leptons}=\sum_l \bar \psi_l \left(i \gamma_\mu \partial^{\mu}-
m_l\right)\psi_l,
\end{equation}
${\cal L}$ is given in eq. (\ref{lag1}) and $l$ describes the
two lightest leptons, i.e., the electron and the muon, whose masses are
respectively $m_e=0.511$ MeV and $m_{\mu}=106.55$ MeV. The
expressions for the energy density ${\cal E}_{lb}$ and the pressure
${P}_{lb}$ are also modified by the leptons, reading:
\begin{equation}
{\cal E}_{lb}={\cal E} +
2 \sum_l \int \frac{d^3p}{{(2 \pi)}^3} \sqrt{{\mathbf p}^2+m_l^2}
(f_{l+} + f_{l-})
\label{enerlep}
\end{equation}
with
\begin{equation}
f_{l\pm} \,=\, \frac{1}{1+\exp[(\epsilon\mp\mu_l)/T]}\;, \quad l=e,\mu ,
\end{equation}
where $\mu_l$ being the chemical potentials for leptons of type $l$,
$\epsilon=\sqrt{{\mathbf p}^2+m_l^2}$ and
\begin{equation}
{P}_{lb}=P + \frac{1}{3 \pi^2} \sum_l \int \frac{{\mathbf p^4} dp}
{\sqrt{{\mathbf p}^2+m_l^2}} (f_{l+} + f_{l-})
\label{presslep}
\end{equation}
where
${\cal E}$ and $P$ are given by eqs. (\ref{enerd}) and
(\ref{pressure}) respectively.

Notice that the leptons are considered as a gas of non-interacting
relativistic particles, in such a way that the minimization of
the thermodynamic potential is not altered by their presence.
The already mentioned requirement of charge neutrality yields
\begin{equation}
\rho_p=\rho_e + \rho_{\mu},
\end{equation}
where the electron and muon densities can be read off from equation
(\ref{rhoi})
by substituting $i$ by $l$. From the condition of chemical equilibrium in the
weak processes, obtained
from the minimization of the Gibbs potential with the conditions of baryon
number and electric charge conservation, one is left with the following
relations between the chemical potentials
\begin{eqnarray}
\mu_p&=& \mu_n- \mu_e\;, \\
\mu_{\mu} &=& \mu_e.
\end{eqnarray}
Common definitions for the lepton fractions are
$Y_l = \rho_l/\rho$, although the lepton densities are not part of
the baryon density. Some consequences of the imposition of $\beta$ stability
in relativistic models are discussed in \cite{em}.

\section{Results and Conclusions}

\begin{figure}
\begin{center}
\epsfig{file=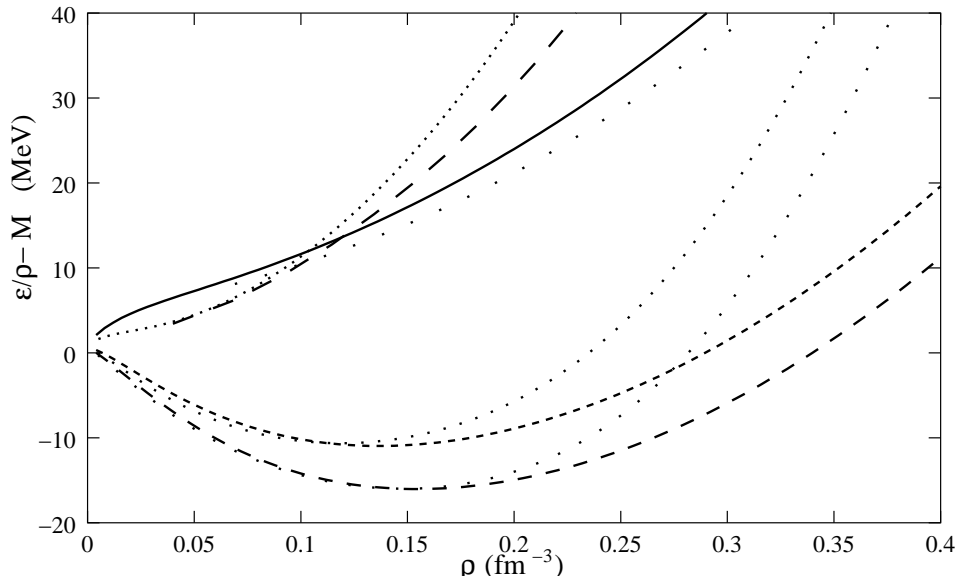}
\caption{Binding energy in terms of the baryon density for
different proton fractions and $T=0$. From top to bottom we show
the EOS with $\beta$-stability for NL3 (dotted line) and for TW
(long-dashed line); $Y_p=0$ for NL3
(solid line) and TW (large spaced dotted line);
$Y_p=0.3$ for NL3 and TW and $Y_p=0.5$ for NL3 and TW.} \label{fig1}
\end{center}
\end{figure}

\begin{figure}
\begin{center}
\epsfig{file=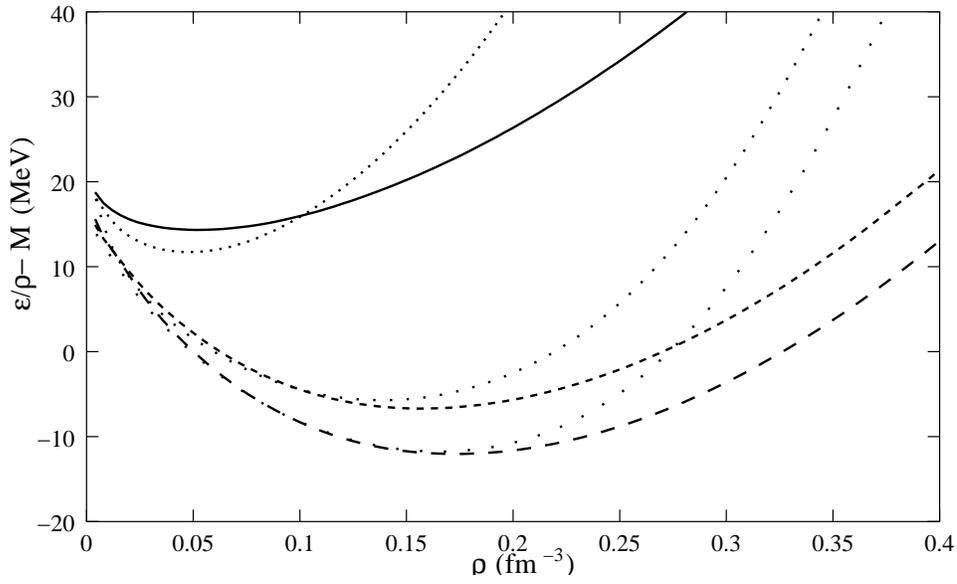}
\caption{Binding energy in terms of the baryon density for
different proton fractions and $T=10$ MeV. From top to bottom we show the EOS
with $Y_p=0$ for NL3 and TW; $Y_p=0.3$ for NL3 and TW; $Y_p=0.5$ for NL3
and TW.}
\label{fig2}
\end{center}
\end{figure}

In figure \ref{fig1} we show the zero temperature EOS for different proton
fractions and two of the parameter sets used in this work, i.e., NL3 and TW.
The TW
parametrization makes the EOS softer not only for symmetric nuclear matter
($Y_p=0.5$), as discussed in \cite{tw}, but also for all other proton fraction
possibilities. The same is true if $\beta$ stability is imposed.
In figure \ref{fig2} the EOS is plotted for $T=10$ MeV and again, a
behavior similar to that of figure \ref{fig1} is observed.
Notice, however, that the minima of all curves are shifted upwards and that
the curves for $Y_p=0$, which do not exhibit minima for $T=0$ acquire them
once the temperature increases.

\begin{figure}
\begin{center}
\epsfig{file=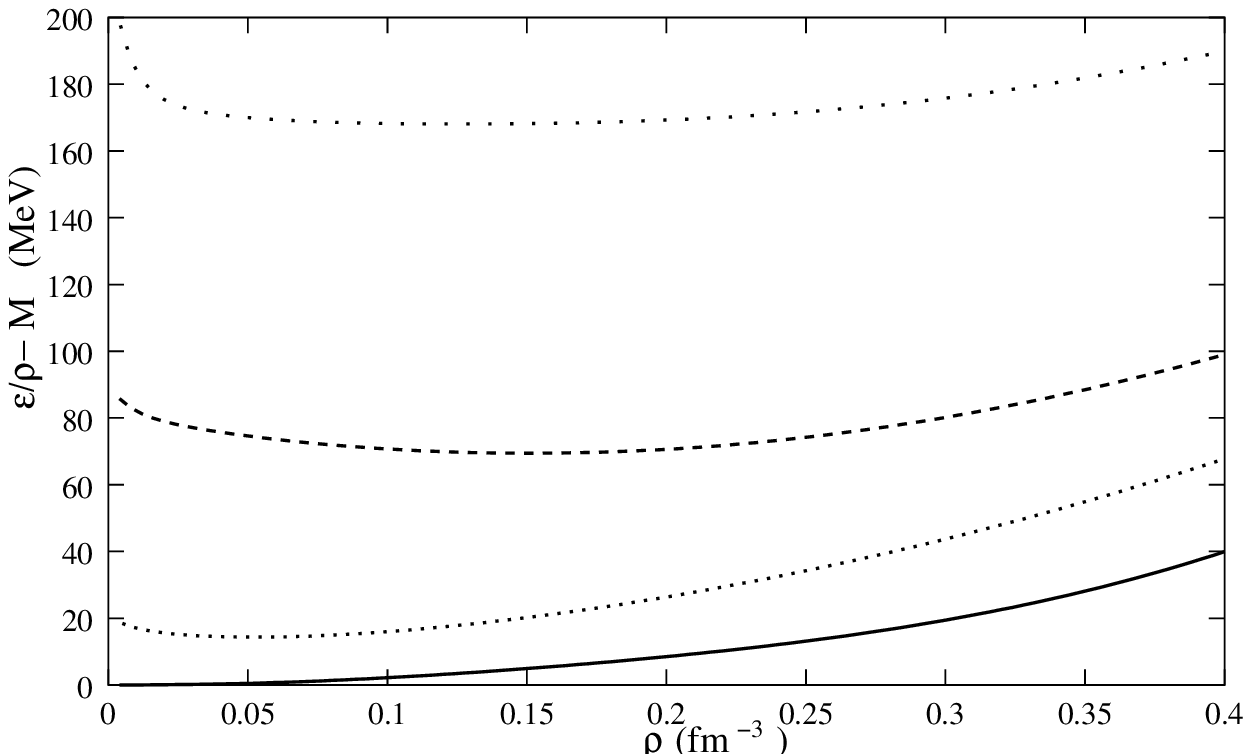}
\caption{Binding energy in terms of the baryon density for
different temperatures, $Y_p=0$ and TW. From top to bottom we show the EOS
for $T=100$ MeV, $T=50$ MeV, $T=10$ MeV and $T=0$.}
\label{fig3}
\end{center}
\end{figure}

\begin{figure}
\begin{center}
\epsfig{file=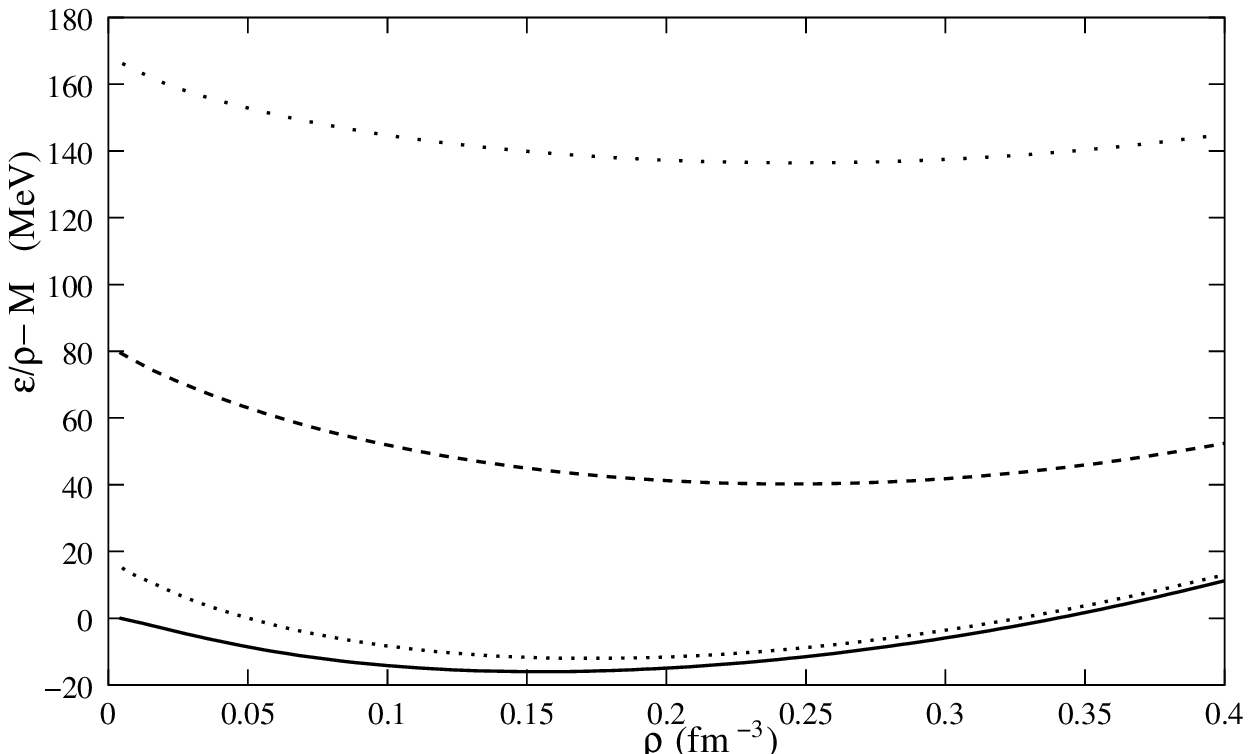}
\caption{Binding energy in terms of the baryon density for
different temperatures,  $Y_p=0.5$ and TW. From top to bottom we show the EOS
for $T=100$ MeV, $T=50$ MeV, $T=10$ MeV and $T=0$.}
\label{fig4}
\end{center}
\end{figure}

In figure \ref{fig3} we show the EOS for neutron matter ($Y_p=0$)
at different temperatures, namely, $T=0$, $T=10$ MeV, $T=50$ MeV
and $T=100$ MeV. At very low densities the inclination of the
curves vary substancially from low to high temperatures. This is
because in this region of low densities, the thermal energy $kT$
is an appreciable fraction of the Fermy energy $\epsilon_F$,
making the effects of the temperature, in particular the
particle-antiparticle creation, more dramatic in this regime
than at high densities, were $\epsilon_F$ is greater.
In figure \ref{fig4} the EOS is plotted, this time for symmetric nuclear
matter. One can see the change in the minumum from a negative to
a positive value, which becomes very large for high temperatures. Once can
also notice that the minima of all curves are slightly shifted to higer
densities.

\indent From the analysis of figures \ref{fig1}-\ref{fig2} we
conclude that the TW parametrization is softer than the NL3 one.
This can be explained looking for the $\Gamma$ parametrizations
in the limit of $\rho/\rho_{sat} \gg 1$. In this limit we have
\begin{eqnarray}
\Gamma_i(\rho)&\longrightarrow&
\frac{a_ib_i}{c_i}\:\Gamma_i(\rho_{sat})\approx
0.7\:\Gamma_i(\rho_{sat})\;,\quad i=\sigma,\omega\label{lim1}
\\
\Gamma_{\rho}(\rho)&\longrightarrow&0 .\label{lim2}
\end{eqnarray}
At such high densities, the system interacts mainly trough the exchange
of the meson $\omega$, once the scalar meson $\sigma$ saturates as
$m^*\longrightarrow 0$. The $\Gamma_\omega$ coupling constant of the NL3 model
is the same as at the saturation density, while equation (\ref{lim1}) says
that, in this limit, the
$\Gamma_\omega$ coupling constant for the TW parametrization is lower than
the value at the saturation point. TW is thus less repulsive at
high densities than NL3, which makes its EOS softer. This fact has
important consequences, for example when modeling neutron stars. A soft
EOS provides a neutron star with a total mass lower than the value obtained
with a stiff EOS.

\begin{figure}
\begin{center}
\epsfig{file=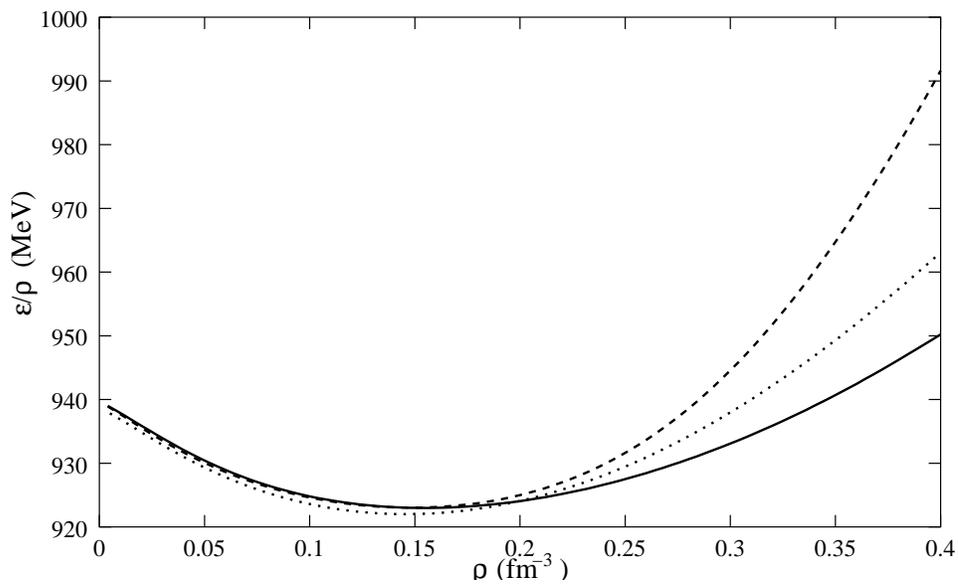}
\caption{Energy density per nucleon in terms of the baryon density for
zero temperature and  $Y_p=0.5$. From top to bottom we show the EOS
for NL3, TM1 and TW.}
\label{fig5}
\end{center}
\end{figure}

\begin{figure}
\begin{center}
\epsfig{file=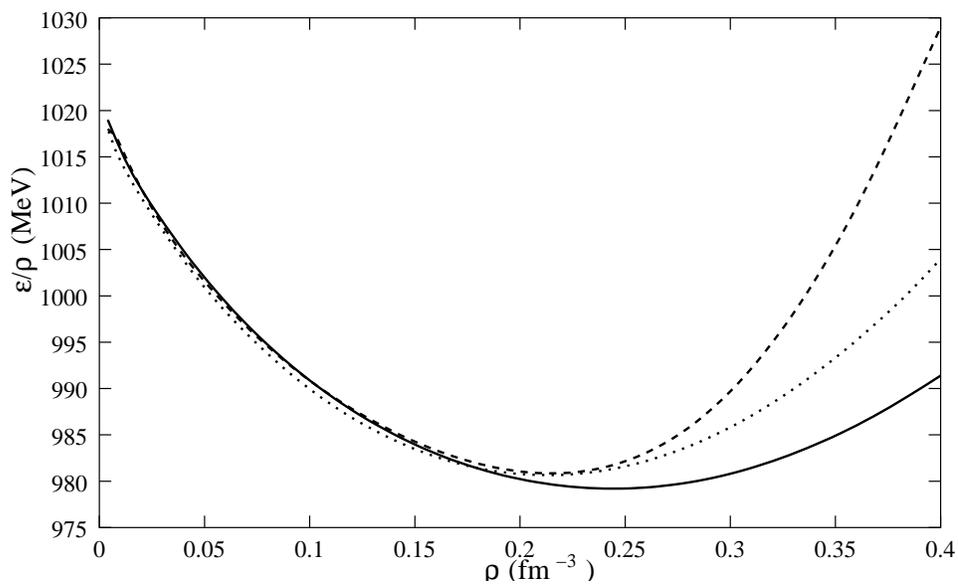}
\caption{Energy density per nucleon in terms of the baryon density for
$T=50$ MeV and  $Y_p=0.5$. From top to bottom we show the EOS
for NL3, TM1 and TW.}
\label{fig6}
\end{center}
\end{figure}

We have also checked that the TW parametrization provides an EOS softer than
the one obtained with the TM1 force \cite{tm1,toki} and closer to the
relativistic Brueckner-Hartree-Fock (RBHF) EOS \cite{bm}, as can be seen in
figures \ref{fig5} and \ref{fig6} for $T=0$ and $T=50$ MeV and symmetric nuclear matter.
The same is true for other proton fractions. The RBHF theory produces well
the nuclear matter saturation based on the nucleon-nucleon interaction
determined by scattering experiments. The TM1 parametrization includes a
non-linear $\omega$ term and hence works with one extra parameter which is
also adjusted in order to reproduce nuclear matter bulk properties. We then
conclude that the TW parameter set is a very useful force in the studies
involving EOS at high densities.

\begin{figure}
\begin{center}
\epsfig{file=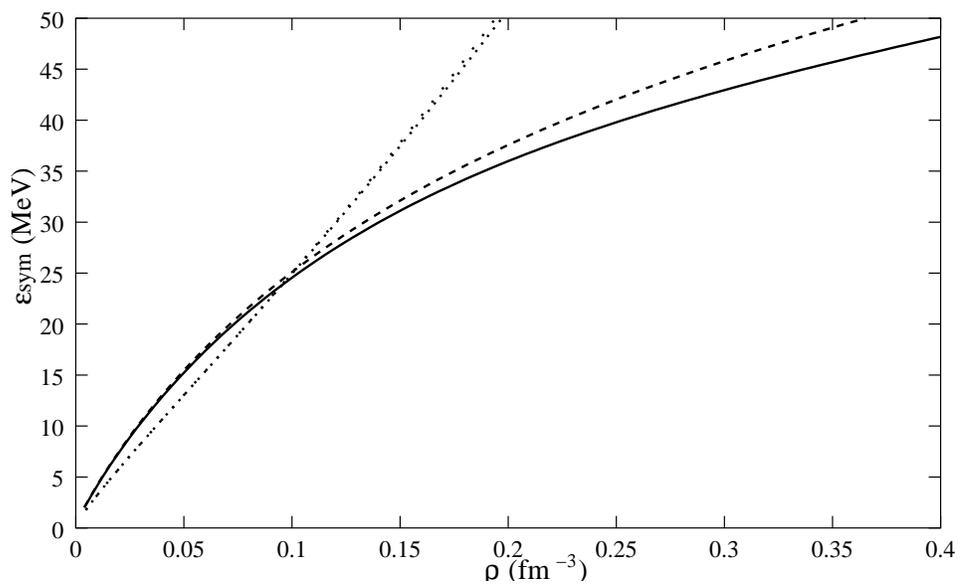}
\caption{Symmetry energy in terms of the baryon density respectively for
$Y_p=0.0$ and $0.5$ for NL3 (dotted curves) and TW (dashed and solid curves).}
\label{fig7}
\end{center}
\end{figure}

In figure \ref{fig7} the symmetry energy is displayed for NL3 and TW for
pure neutron matter and symmetric nuclear matter. Different proton fractions
give rise to slightly different curves because of the difference in Fermi
momenta and in the effective mass, which enters in $E^{\ast}(\mathbf p)$.
According to \cite{lkb}
the symmetry energy at normal nuclear matter density is found to lay
in between 27 and 36 MeV in the mass formula calculations, in the range of
28 to 38 MeV in non-relativistic models and in between 35 and 42 MeV in
relativistic models. Notice that at the saturation
point, the value for the TW parametrization (32 MeV) is somewhat lower than
for the NL3, remains in the accepted range of validity and is
closer to the predictions of non-relativistic models. Moreover,
the curves
obtained for the TW model present a much smaller symmetry energy at larger
densities and also a smoother behavior as compared with the curves arising
from the NL3 model, which gives a more linear tendency to the curve.
This result can be explained looking at equation
(\ref{lim2}), which tell us that $\Gamma_{\rho}$ goes to
zero at high densities, consequently so does its contribution in equation
(\ref{esym}). On the opposite, $\Gamma_{\rho}$ in the NL3
parametrization is constant as a function of density.

\begin{figure}
\begin{center}
\epsfig{file=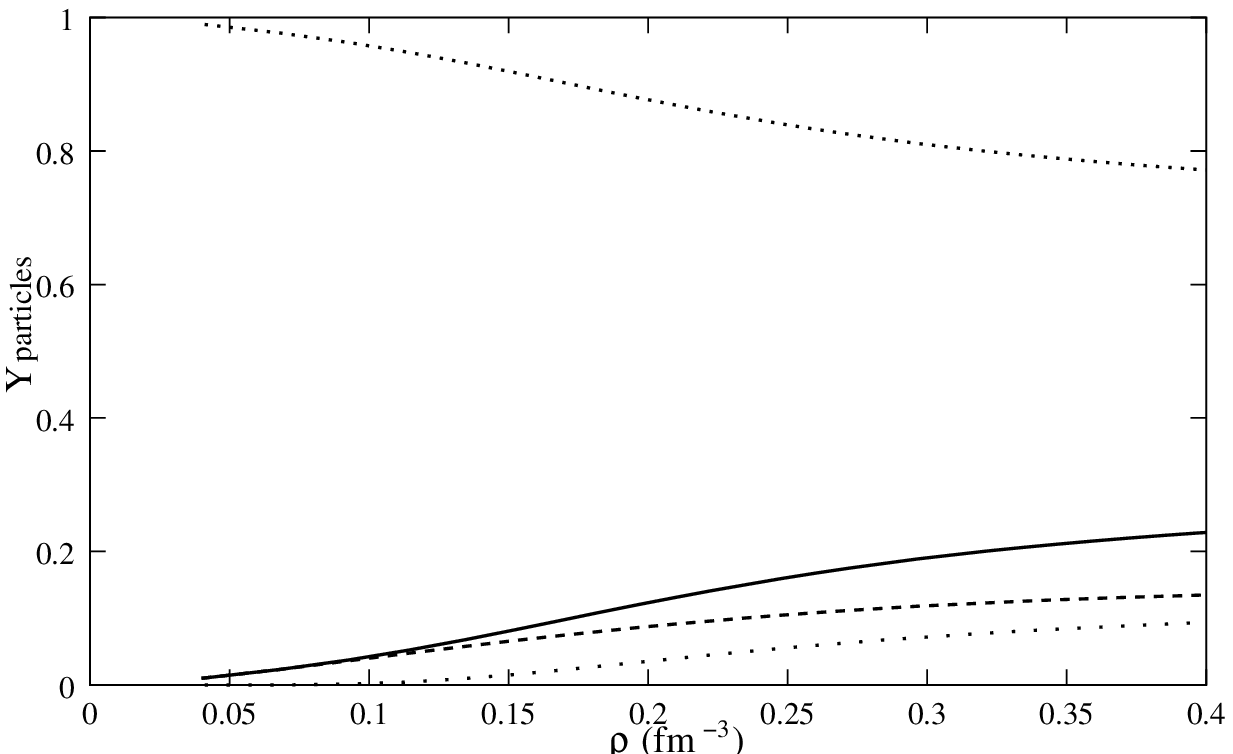}
\caption{Particle composition in terms of the baryon density for
$T=0$ and NL3. From top to bottom we show the distribution of neutrons,
protons, electrons and muons.}
\label{fig8}
\end{center}
\end{figure}

\begin{figure}
\begin{center}
\epsfig{file=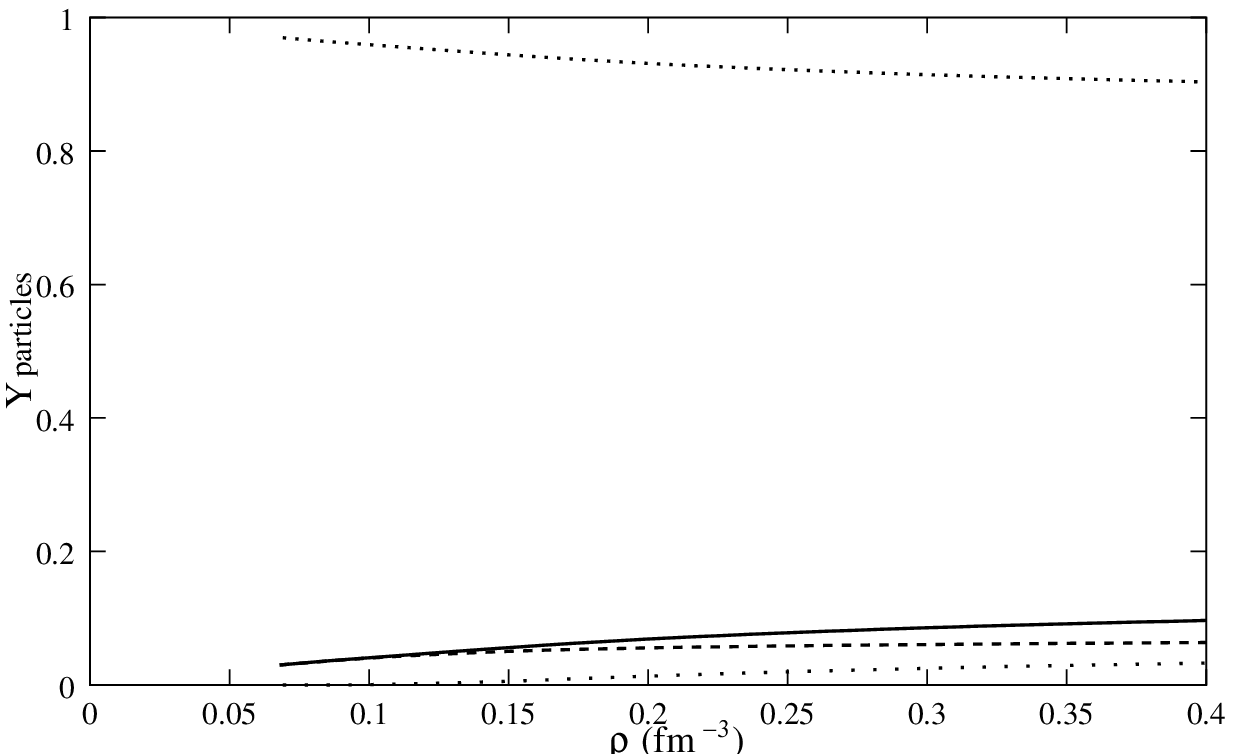}
\caption{Particle composition in terms of the baryon density for
$T=0$ and TW. From top to bottom we show the distribution of neutrons,
protons, electrons and muons.}

\label{fig9}
\end{center}
\end{figure}

\begin{figure}
\begin{center}
\epsfig{file=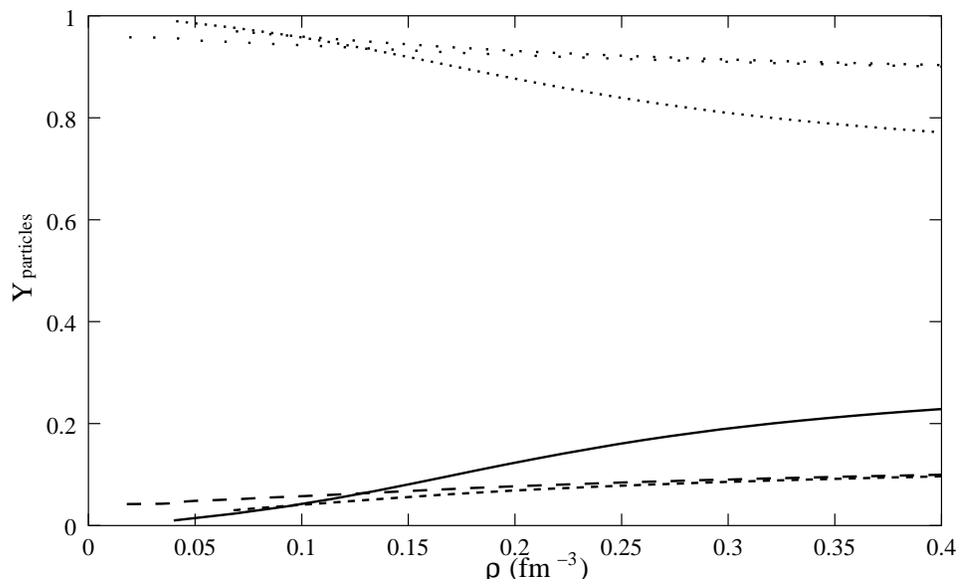}
\caption{Proton and neutron composition in terms of the baryon density.
From top to bottom, looking at the lefthand side of the figure
we show the distribution of neutrons for NL3 and $T=0$, TW and $T=0$, TW and
$T=10$ MeV, and for protons for TW and $T=10$ MeV, TW and $T=0$ and NL3 and
$T=0$.} \label{fig10}
\end{center}
\end{figure}

We have finally studied the particle composition once $\beta$
stability is imposed. In figures \ref{fig8} and \ref{fig9} the particle
composition obtained at $T=0$ respectively for NL3 and TW are shown. In
figure \ref{fig10} we show the  proton and neutron composition for NL3 and
TW at $T=0$ MeV and TW at $T=10$ MeV. One can see
that if the temperature does not increase much, the particle
composition for a fixed parameter set remains basically the same.
Nevertheless, it changes substancially from NL3 to TW. In
particular, at high densities, we can see that TW is less
isospin-symmetric than NL3. This is due, again, to the result
(\ref{lim2}), which says that, for TW, the $\rho$-nucleon
interaction is suppressed at high densities. It is precisely this
interaction which drives the systems to a more isospin-symmetric
configuration at high densities, as we can see in figure
(\ref{fig8}) for NL3, where this interaction survives in this
limit.

An extension of this work in order to study liquid-gas phase transition
and consequent droplet formation is currently under investigation.

\section*{ACKNOWLEDGEMENTS}
This work was partially supported by CNPq - Brazil.
One of the authors (D.P.M.) would like to thank Dr. Constan\c ca Provid\^encia
for very useful suggestions and productive discussions related to this work.

%\newpage

\begin{table}
\centerline{{\bf Table 1} Nuclear matter properties.}
\vspace{0.5cm}
\begin{center}
\begin{tabular}{|c|c|c|c|c|c|c|c|c|c|c|}
\hline
   &  NL3 \cite{nl3}  &  TM1 \cite{tm1} & TW \cite{tw} \\
\hline
$B/A$ (MeV) & 16.3 & 16.3 & 16.3\\
\hline
$\rho_0$ (fm$^{-3}$) & 0.148 & 0.145 & 0.148 \\
\hline
$K$ (MeV) & 272 & 281 & 240 \\
\hline
${\cal E}_{sym.}$ (MeV)  & 37.4 & 36.9 & 32.0 \\
\hline
$M^*/M$ & 0.60 & 0.63 & 0.56\\
\hline
\end{tabular}
\end{center}
\end{table}

\end{document}